\documentclass[aps, prd, twocolumn, showpacs, superscriptaddress, groupedaddress]{revtex4} 

\usepackage{graphicx}  
\usepackage{dcolumn}  
\usepackage{bm}        
\usepackage{amssymb}  
\usepackage{color}

\usepackage[pagebackref=false, colorlinks=true]{hyperref}
\hypersetup{
linkcolor=blue,     
citecolor=blue,     
urlcolor=blue}      

\begin{document}

\title{A General Relativistic Approach to Small-Scale Structure Formation}

\author{Dipanjan Dey}
\email{dipanjandey.adm@charusat.edu.in}
\affiliation{International Center for Cosmology, Charusat University, Anand 388421, Gujarat, India}
\author{Prashant Kocherlakota}
\email{k.prashant@tifr.res.in}
\affiliation{Tata Institute of Fundamental Research, Mumbai 400005, India}
\author{Pankaj S. Joshi}
\email{psjprovost@charusat.ac.in}
\affiliation{International Center for Cosmology, Charusat University, Anand 388421, Gujarat, India}

\date{\today}

\begin{abstract}
We treat here general relativistically the issue of galaxy formation, which is a major problem in cosmology. While the current models use a top-hat collapse model, coupled with Newtonian virialization technique to balance the gravitationally collapsing matter cloud into a galaxy, we present here a general relativistic toy model to achieve such a purpose. 
We consider a relativistic gravitational collapse that begins from physically reasonable and non-singular initial conditions and that tends to an equilibrium configuration in asymptotic time. The matching of different spacetime regions is explicitly demonstrated to establish the feasibility of the model. This helps us understand better how the formation of galaxy-like objects and dark matter halos are likely to develop as the universe evolves, using a general relativistic technique. While the toy model we present here uses a somewhat simplistic collapse framework, this, however, has the potential to develop into a more realistic scenario, as we have noted. As we point out, 
the attractive feature here is, we have explicitly demonstrated that equilibrium configurations can be formed, with suitable matchings made, which go some way to treat the problem of galaxy formation within a full general relativity framework. 
\end{abstract}

\pacs{}
\maketitle

\section{Introduction}
Cosmological structure formation is one of the most interesting topics at the forefront of cosmology and astrophysics today. It deals with questions such as, how do dark matter structures and galaxies form? The current consensus is that quantum fluctuations in the inflationary era are responsible for the large scale structure of our present universe \cite{LiddleLyth93, white, Sengor18}. These quantum fluctuations create matter density perturbations, which, at the conclusion of the era of radiation domination, produce a local density contrast, $\delta\rho/\bar{\rho} = \rho/\bar{\rho} - 1$, with $\bar{\rho}$ representing background density. The dynamics of these density perturbations is initially tackled within linear perturbation theory, for as long as the density contrast remains much smaller than one. With time, these perturbation modes grow and gradually enter the non-linear regime, $\delta\rho/\bar{\rho}\sim 1$, resulting in various `over-dense' patches in the universe.  Naturally, one is then compelled to use non-linear techniques to describe the dynamics of these non-linear perturbation modes. As is well known, dark matter decouples from the `primordial soup' far before baryonic matter and starts collapsing to form structures, and thus these over-dense regions are typically thought to contain mostly dark matter. As a first approximation, these over-dense patches are taken to be spherically symmetric and the evolution of these regions is generally modeled by the `top-hat collapse model' \cite{GunnGott72}. In this model, these over-dense regions are described via a spherically symmetric closed Friedmann-Lemaitre-Robertson-Walker (FLRW) metric. On the cosmological scale, the universe is almost flat, and therefore the background of the over-dense regions is described by spatially flat FLRW metric.

According to the top-hat collapse model, over-dense regions expand initially in an isotropic, homogeneous fashion along with the `background' flat FLRW spacetime. The fluid of the spherically symmetric over-dense regions is considered to be a homogeneous, pressure-less fluid. Eventually, the dynamics of these over-dense regions detaches from the background cosmic expansion and these regions start to behave like sub-universes. These over-dense sub-universes start to collapse under gravity, with the matter content there modeled as remaining homogeneous, isotropic and pressure-less throughout the collapse. Now, within general relativity (GR), it is well known that the final fate of a collapse involving a homogeneous pressure-less fluid (dust) is always a black-hole \cite{OppenheimerSnyder39, Datt, Joshi07}, and so via this top-hat collapse model, one cannot relativistically obtain desired small-scale equilibrium configurations as eventual end-states of the gravitational collapse of over-dense regions. Consequently, \textit{Newtonian} virialization techniques are used to obtain such an equilibrium, to explain galaxy formation. 

Virialization is a process in which a system of $N$ particles achieves an equilibrium state at a large system time when,
\begin{equation} \label{eq:vir1}
\langle T\rangle_{\tau\rightarrow\infty} = -\frac12\Bigg\langle \sum_{i=1}^{N}{\bf F}_i\cdot{\bf r}_i  \Bigg\rangle_{\tau\rightarrow\infty},
\end{equation}
where ${\bf F}_i$ is net force on the $i$th particle, ${\bf r}_i$ is its position, $T$ is the total kinetic energy of the system and the system reaches the virialized state in a large system time $\tau$. The angular bracket denotes time averaging. If the particles of such a system interact only gravitationally with each other, then the system reaches its final equilibrium state when the following condition is fulfilled,
\begin{equation}
\langle T\rangle=-\frac{1}{2} \langle V_T \rangle,
\end{equation}
where $V_T$ is the total gravitational potential of the system. There are primarily four processes governing a collapsing system, consisting only of gravitationally interacting particles, which achieves a virialized state: violent relaxation, phase mixing, chaotic mixing and Landau damping \cite{Lynden-Bell67, Merritt99}. Generally, a collapsing system virializes in very large system time. However, since these processes are all Newtonian, one is uncertain about the causal structure of the final stable system.

In the top-hat collapse model, the above mentioned virialization argument is invoked to stabilize a collapsing system. In this model, dark matter is considered to be homogeneous and dust-like throughout the evolution of the over-dense regions, the main reason behind this being that such a fluid can satisfactorily explain the large-scale structure of our universe. Models in which dark matter is considered to be pressureless and non-relativistic are known as the cold dark matter (CDM) models \cite{BullockBoylan-Kolchin17, Weinberg+15}. CDM model is very successful in describing cosmological large scale structure. However, it is unsatisfactory in providing answers to various problems regarding small scale structures such as galaxies. To resolve such problems involving dark matter, self-interaction between dark matter particles is considered \cite{SpergelSteinhardt00}. However, all of these models rely on this unsatisfactory and a bit \textit{ad-hoc} virialization technique, to attain the final equilibrium state of a collapsing system. Another problem with these models is that during collapse the matter field is assumed to remain homogeneous, which is also unrealistic. 

As alluded to before, dark matter decouples first and starts forming over-dense structures. Eventually, baryonic matter starts to fall down the gravitational potential created by these over-dense dark matter pockets. Now, since dark matter interacts only via gravity, it cannot radiate away its energy via electromagnetic waves, etc, and eventually forms large halo-like structures. The baryonic matter, however, can radiate away its energy and is capable of collapsing further to form stable structures that are smaller in size than the dark matter halos. Therefore, there is an accumulation of baryonic matter in the central region of these spherical, over-dense dark matter halos, and the density of baryonic matter is far smaller in the outer regions of the halo. Outside this central region of the dark matter halo, the gravity of dark matter dominates and local tidal forces in the halo prevent the collapse of baryonic matter to form stars. However, as baryonic matter gradually accumulates in the central region of the halo, the possibility of star formation there increases. As baryonic matter collapses into the central region, the rotation speed of accreting matter increases and at some time, the centrifugal force due to this rotation balances the gravitational force. In this way, galaxies are born inside a massive dark matter halo \cite{Padmanabhan93, FrenkWhite12, CooraySheth02}. 


Our purpose here is to address these two problems, namely that  
of ad-hoc virialization and homogeneous collapse, by implementing a general relativistic technique to attain equilibrium from collapse, following \cite{JMN11}. The aim is to describe baryonic structure formation inside a massive dark matter halo and we ignore rotation in the baryonic matter here for simplicity. Our model of small scale structure formation is valid until the rotation speed of baryonic matter becomes significantly high. We begin in \S\ref{sec:StructureFormation} with a brief review of the top-hat collapse model and conventional theory of galaxy formation. In \S\ref{sec:BMDM}, we then proceed to discuss our model for cosmological baryonic matter (BM) and dark matter (DM) content as two non-interacting perfect fluids, and how this two-component perfect fluid can be described effectively using a single \textit{anisotropic} fluid. We will then make a simplifying assumption regarding this effective anisotropic fluid: we take its radial pressure to be vanishingly small. This is done to link up with the already existing Joshi-Malafarina-Narayan-1 (JMN-1) spacetimes which describe the geometry of equilibrium configurations that form as asymptotic end-states of a general relativistic gravitational collapse process \cite{JMN11}. Additionally, the JMN-1 spacetimes can have, at asymptotic times, arbitrarily large energy densities near the center of the spacetime, a useful feature to model the ultra-dense regions thought to be present at the centers of galaxies. One can theoretically predict what would be the observational consequences if there exists an ultra-high density region at the center of a galaxy like structure \cite{Shaikh:2018lcc, Chakraborty:2016mhx, Dey:2013yga, Dey+15}. In  \S\ref{sec:JMN1}, we set up this collapse process and discuss the JMN-1 spacetime and its properties in some detail. In \S\ref{sec:Galaxy_Halo}, we present a new general relativistic model of small scale structure formation. We begin with a discussion on how over-dense regions form. Then, we use suitable matching conditions to explain different epochs of formation of a galaxy like structure and dark matter halo. Finally, we end with a discussion and summary of our results in \S\ref{sec:Discussion}.

\section{Galaxy and Dark Matter Halo Formation}\label{sec:StructureFormation}

In the following, we first discuss here in some detail the top-hat collapse model towards the halo formation.

\subsection{Halo formation: The Top Hat Collapse Model} \label{sec:TopHat}
In the standard spherical top-hat collapse model, the metric of the overdense region is given by the closed FLRW metric :
\begin{eqnarray} \label{eq:FLRWMetric}
 ds^2 = - dt^2 + {a^2(t)\over 1- r^2 }dr^2 + a^2(t) r^2 d\Omega^2,
\end{eqnarray}
where $a(t)$ is the scale-factor of the closed FLRW spacetime and $r$ ranges from zero to one. As we have discussed, in top-hat collapse the matter of overdense regions is considered  as dust-like. Therefore, the above spacetime is a particular form of the Lemaitre-Tolman-Bondi (LTB) metric. From the Einstein equations we get the following differential equation,   
\begin{eqnarray} 
\frac{H^2}{H_0^2}=\Omega_{m0}\left(\frac{a_0}{a}\right)^3 + (1-\Omega_{m0})\left(\frac{a_0}{a}\right)^2 \, ,
\label{fried}
\end{eqnarray}
where the Hubble parameter for the overdense
sub-universe $H=\dot{a}/a$ and $H_0,\,a_0$ are  the initial values of $H$ and $a$ and the initial values correspond to the time when the dynamics of overdense regions become independent of back-ground expansion.
The above differential equation is known as Friedmann equation. In the above equation, $\Omega_{m0}=\rho_0/\rho_{c0}$, where $\rho_{c0} =
3H_0^2$, and $\rho_0$ is the initial homogeneous matter density.
The solution of the Friedmann equation can be written in a parametric form:
\begin{eqnarray}
 a = \frac{a_m}{2}(1- \cos\theta),~~~
 t = \frac{t_m}{\pi}(\theta-\sin\theta)\, ,
\label{at}
\end{eqnarray}
where $t_m$ is the time when the scale factor $a(t)$ of closed FLRW universe reaches its maximum value $a_m$. The scale factor $a(t)$ can attain its maximum value $a_m$ when $\theta=\pi$. One can always write $a_m$, $t_m$ in terms of $\Omega_{m0}$, $H_0$ and $a_0$ as,
$$a_{m}={a_0 \Omega_{m0}\over (\Omega_{m0}-1)}\,,\,\,\,\, t_{m}={\pi \Omega_{m0}\over 2 H_0 (\Omega_{m0}-1)^{3/2}}\,,$$
where $\Omega_{m0}>1$ due to the over-density. As in top-hat collapse model the overdense regions are considered as sub-universes, these sub-universes only follow the dynamics of closed FLRW metric. These overdense sub-universes expanded first in a homogeneous and isotropic manner and at some turnaround time $t=t_m$ those regions stopped expanding and started to collapse under its own gravity. At the turnaround point $t=t_m$ we can always calculate the ratio between the density of overdense region and the density of background:
\begin{eqnarray}
\frac{\rho(t_{m})}{\bar{\rho}(t_{m})} = \frac{9\pi^2}{16} \sim 5.55\,, 
\label{impfac}
\end{eqnarray}
which indicates that at turnaround point the overdense regions are $5.55$ times denser than the back-ground. In Top-Hat collapse model as the matter field is considered as homogeneous and dust-like throughout the collapse, the final state of collapse is always a black hole. The final singularity at $\theta= 2\pi$ must be covered by event horizon. To avoid this total collapse, virialization technique is used to stabilize the collapsing system. We know that a self-gravitating system virializes when the following condition is satisfied,
\begin{equation}
\langle T \rangle=-\frac12 \langle V_T \rangle \, ,
\end{equation}
where $ \langle T \rangle$ and $\langle V_T \rangle$ are the average total kinetic and potential energy respectively.
Now, let us consider an overdense spherical region of mass $M$ that reaches its maximum scale factor limit at the turnaround time. At that time kinetic energy is zero, and all the energy is potential energy,
\begin{equation}
V_T=-\frac{3M^2}{5R_m}\, ,
\end{equation}
where $R_m$ is the maximum physical radius of the overdense sub-universe at the turnaround time $t_m$. When it has collapsed to half of its maximum physical radius then from the conservation of total energy, we can derive the expressions of potential energy and kinetic energy,
\begin{eqnarray}
V_T&=& -\frac{6M^2}{5R_m}\, ,\nonumber\\
T&=& \frac{3M^2}{5R_m} = -\frac{V_T}{2}\, .
\end{eqnarray}
Since at this point total kinetic energy is half of absolute value of total potential energy, the system virializes to a stable configuration. 
So, in top-hat collapse model the collapsing overdense regions reach the virialized state when
$a_{\rm vir}=\frac12 a_{m}$. Using  eq.~(\ref{at}), one can show that in this model the system virializes when $\theta = \frac{3\pi}{2}$.  At this stage the ratio between the density of overdense region and the density of back-ground becomes,
\begin{eqnarray}
  \frac{\rho(t_{\rm vir})}{\bar{\rho}(t_{\rm vir})} \sim 145\,,
\label{ratio2}
\end{eqnarray}
which implies that at the time $t_{\rm vir}$ when  dark matter virializes to its stable halo structure, the density of the halo becomes $145$ times larger than background. These numbers which come out from the spherical Top-Hat collapse model are very important for astrophysics.

There are some drawbacks, however, in the spherical Top-Hat collapse model, though it gives some important predictions. 
In this model the matter fluid is always homogeneous, dust-like. The overdense regions in this model start to collapse with homogeneous matter field and remain homogeneous throughout the collapse which is unrealistic. The second problem with this model is that in this model, to describe the total non-linear evolution of overdense regions general relativistic technique is used. However, to describe the equilibrium configuration of collapsing system the Newtonian virialization technique is used. For more realistic scenario, one can introduce in-homogeneous density and non-zero pressure in collapsing matter cloud. There are many literature where collapse of different types of fluid are investigated \cite{Joshi:2012mk}-\cite{JMN11}, \cite{GoswamiJoshi04}.
In this paper we will use a general relativistic technique of equilibrium \cite{JMN11}, which can be thought of as a general relativistic analog of virialization \cite{Dey2+18}.

\subsection{Galaxy Formation}
\label{galform}
The galactic scale fluctuations in dark matter field form the overdense regions of dark matter which expands first, turns around, collapses and virializes to a stable halo-like structure. However, during the collapse process of dark matter halo, gas of baryonic matter starts to collapse inside the halo due to the gravitational potential of halo. As the baryonic matter collapses, gradually, at different phases, different structures form. 
There are mainly two different models for galaxy formation.
In one model \cite{white},\cite{White:1977jf},\cite{Hogan:1985bc},\cite{White:1994bn}, baryonic gas inside the dark matter halo collapses with the halo, heat up     
and virializes at a virialization temperature. The temperature of shocked and pressure supported baryonic gas can be well above the temperature $T\sim 10^4 K$, and therefore the gas becomes ionized. This ionized gas cannot be further pressure supported as it radiates its energy and cools down. One can derive an expression of cooling time scale $t_{cool}$, after which the ionized gas cannot retain its virialized structure.

The cooling rate depends upon the rate of bremsstrahlung, recombination and collisionally excited emission. The dynamics of the gas depends upon the ratio of $t_{cool}$ and Hubble time and also on the ratio of $t_{cool}$ and dynamic time $t_{dyn}$, where the dynamic time is the total evolution time of dark halo. In Top-Hat collapse model $t_{dyn}\sim 1.81t_m$. If the cooling time scale $t_{cool}\lesssim t_{dyn}$, then the major portion of gas inside the halo cools down and collapses to the central region of the halo. As this gas collapses, it creates a very high baryonic matter density around the center. When central gas density dominates over dark matter density around the center, baryonic matter starts the fragmentation process and forms stars. However, this will not explain the flat surface of the spiral galaxy. For that, we need to include rotation. As the central gas collapses, the rotation speed of accreting matter field increases and at some time the central matter field achieves rotational stability.
On the other hand, if $t_{cool}> t_{dyn}$, then the gas continuously has pressure in it which makes the collapse of the gas quasi-static. So, for this case gas slowly collapses with halo and starts fragmentation when the gas becomes self-gravitating system near the central region of halo.

In another model \cite{Binney1},\cite{Birnboim:2003xa}, it is assumed that the major fraction of gas inside the dark-halo cools down before it reaches the virialization temperature and it collapses to the central halo region and starts fragmentation process as it was discussed before.
In some literature \cite{Birnboim:2003xa}, it is shown that  for the first model we need massive halo and virialization of shocked baryonic gas which is not possible in small halo. Therefore, the scenario which is described by the second model is applicable in small halos.

Our attempt here is to present a general relativistic technique, by which we can model the dark matter and baryonic matter structure formation. In this toy model given here, we describe only the collapse and equilibrium process general relativistically, thus approximating the dissipation as a Newtonian process. It should be noted that in our model the range of comoving radius is $0\leq r\leq r_b$, where $r_b$ is the matching radius of an internal metric with an exterior Schwarzschild metric and $r_b<1$. In Top-Hat collapse, the range is $0\leq r\leq 1$.  In \cite{Bhattacharya:2017chr}, we have discussed the final equilibrium configuration in the cosmological scenario, with this close range of $r$.

\section{Modeling Cosmological Baryonic and Dark Matter} \label{sec:BMDM}
In general relativity, perfect fluids have been the most widely used matter models, in both static and dynamical settings. Such models help make dynamics, prescribed via the Einstein equations, significantly more tractable. More generally, when there are multiple fluid components in a system of interest, it is difficult to describe their dynamics using a single effective fluid, due to possible interactions between fluid components. A system of non-interacting perfect fluids, however, is simpler to analyze. Further, such a system comprising multiple non-interacting perfect fluids can be analyzed by their replacement with a single effective anisotropic fluid \cite{Letelier80, Bayin85, Dey+15}. Fluids with anisotropic pressure have been studied in cosmology and astrophysics \cite{Bayin85}. 

Here we shall treat cosmological baryonic matter and dark matter as perfect fluids, and neglecting interactions between them, 
we outline how they can be treated effectively as a single anisotropic fluid. The BM and DM perfect fluids satisfy,
\begin{eqnarray} \label{eq:EoS_BMDM}
p_{\text{B}} &=& \gamma_{\text{B}}\rho_{\text{B}}, \\
p_{\text{D}} &=& \gamma_{\text{D}} \rho_{\text{D}}, \nonumber
\end{eqnarray}
where $\gamma_{\text{B}}$ and $\gamma_{\text{D}}$ are the equation of states of the baryonic and dark matter respectively. Their stress-energy tensors are respectively given as,
\begin{eqnarray}
t^{\mu\nu}_{\text{B}} &=& [(1 + \gamma_{\text{B}})u^\mu u^\nu + \gamma_{\text{B}}g^{\mu\nu}]\rho_{\text{B}}, \\
t^{\mu\nu}_{\text{D}} &=& [(1 + \gamma_{\text{D}})v^\mu v^\nu + \gamma_{\text{D}}g^{\mu\nu}]\rho_{\text{D}}, \nonumber
\end{eqnarray}
Then the stress-energy tensor of the effective anisotropic fluid $T^{\mu\nu}$ formed from the sum of $t^{\mu\nu}_{\text{B}}, t^{\mu\nu}_{\text{D}}$ is,
\begin{eqnarray}
T^{\mu\nu} &=& (1 + \gamma_{\text{B}})\rho_{\text{B}}u^{\mu}u^{\nu} + (1 + \gamma_{\text{D}})\rho_{\text{D}}v^{\mu}v^{\nu} \nonumber \\
&& + (\gamma_{\text{B}}\rho_{\text{B}} + \gamma_{\text{D}}\rho_{\text{D}})g^{\mu\nu}, 
\end{eqnarray}
where $u^{\mu}$ and $v^{\mu}$ are the four-velocities of the two perfect fluids. Most generally, $u^{\mu}$ and $v^{\mu}$ satisfy,
\begin{eqnarray}
u^\mu u_\mu = v^\mu v_\mu = -1, \\
u^\mu v_\mu = \zeta, 
\label{eq:rel_vel_BMBM}
\end{eqnarray}
and when the free function $\zeta$ measures the local relative four-velocity of the BM with respect to the DM, for example, when $\zeta$ vanishes, the four-velocities are orthogonal to each other. Here we take the view that $\zeta$ may be chosen at our convenience since there does not exist an obvious or `natural' physical (or even empirical) constraint on the relative four-velocities of evolving BM and DM in a cosmological setting. %
Transformations that leave the energy-momentum tensor invariant are %
\begin{eqnarray}
u^\mu &\rightarrow & \bar{u}^\mu = \cos\alpha~u^\mu - \Xi~\sin\alpha~v^\mu, \\
v^\mu &\rightarrow &\bar{v}^\mu = \cos\alpha~v^\mu + \Xi^{-1}\sin\alpha~u^\mu,
\end{eqnarray}
where we have defined,
\begin{equation} \label{eq:Xi}
\Xi  =\sqrt{\frac{1 + \gamma_{\text{D}}}{1 + \gamma_{\text{B}}}\frac{\rho_{\text{D}}}{\rho_{\text{B}}}}.
\end{equation}
The specific transformation that makes the transformed four-velocities orthogonal to each other, i.e. $\bar{u}^{\mu}\bar{v}_{\mu}=0$, is obtained by setting $\alpha$ as follows,
\begin{equation} \label{eq:alpha}
\tan(2\alpha) = \frac{2 \zeta \Xi}{1 - \Xi^2}.
\end{equation}
Let us now introduce the normalized fluid velocities,
\begin{equation} \label{eq:normalised_barred_vel}
w^{\mu} = \frac{\bar{u}^{\mu}}{\sqrt{-\bar{u}^\nu \bar{u}_\nu}},\
y^{\mu} = \frac{\bar{v}^{\mu}}{\sqrt{\bar{v}^\nu\bar{v}_\nu}},
\end{equation}
which satisfy $w^{\mu}w_{\mu}=-1$, $y^{\mu}y_{\mu}=1$, and $w^{\mu}y_{\mu}=0$. Note that $w^{\mu}$ is a timelike four-vector and $y^{\mu}$ is \textit{spacelike}. Written in terms of $w^{\mu}$ and $y^{\mu}$, the effective energy-momentum tensor $T^{\mu\nu}$ becomes,
\begin{equation}
T^{\mu\nu} = (\rho + p_\theta)w^{\mu}w^{\nu} + p_\theta g^{\mu\nu} + (p_r - p_\theta)y^{\mu}y^{\nu},
\end{equation}
where we have introduced $\rho, p_r$ and $p_\theta$, the effective energy-density, radial and azimuthal pressures respectively, given via (see \S III.B of \cite{Dey+15})
\begin{widetext}
\begin{eqnarray} \label{eq:Tmunu_effective}
\rho &=& T^{\mu\nu}w_{\mu}w_{\nu} \label{eq:rho_eff}  \\
&=& - \frac{(1 \! - \! \gamma_{\text{BM}})\rho_{\text{BM}} \! + \! (1 \! - \! \gamma_{\text{DM}})\rho_{\text{DM}}}{2} \! + \! \frac{\sqrt{
[(1 \! + \! \gamma_{\text{BM}})\rho_{\text{BM}} \! + \! (1 \! + \! \gamma_{\text{DM}})\rho_{\text{DM}}]^2 + 4(K^2 \! - \! 1)(1 \! + \! \gamma_{\text{BM}})(1 \! + \! \gamma_{\text{DM}})\rho_{\text{BM}}\rho_{\text{DM}}}}{2}, \nonumber \\
p_r &=&  T^{\mu\nu}y_{\mu}y_{\nu} \\
&=& \frac{(1 \! - \! \gamma_{\text{BM}})\rho_{\text{BM}} \! + \! (1 \! - \! \gamma_{\text{DM}})\rho_{\text{DM}}}{2} \! + \! \frac{\sqrt{
[(1 \! + \! \gamma_{\text{BM}})\rho_{\text{BM}} \! - \! (1 \! + \! \gamma_{\text{DM}})\rho_{\text{DM}}]^2 + 4 K^2(1 \! + \! \gamma_{\text{BM}})(1 \! + \! \gamma_{\text{DM}})\rho_{\text{BM}}\rho_{\text{DM}}}}{2}, \nonumber \\
p_\theta &=& \gamma_{\text{BM}}\rho_{\text{BM}} + \gamma_{\text{DM}}\rho_{\text{DM}}.
\end{eqnarray}
\end{widetext}
In comoving coordinates we may choose
\begin{eqnarray}
& w^1 = w^2 = w^3 = 0,\ w^0w_0 = -1, \\
& y^0 = y^2 = y^3 = 0,\ y^1y_1 = 1. \nonumber
\end{eqnarray}
Therefore in these coordinates, i.e. in the proper frame of the effective anisotropic fluid, we can write,
\begin{equation}
T^0_{\ 0} = -\rho,\ T^1_{\ 1} = p_r,\ T^2_{\ 2} = T^3_{\ 3} = p_\theta.
\end{equation}
We can invert the relations given above between the energy-densities  of the individual perfect fluids and the properties of the effective anisotropic fluid to write,
\begin{eqnarray} \label{eq:rho_p_eff}
\rho_{\text{BM}} &=& \frac{\gamma_{\text{DM}}(\rho - p_r + p_\theta) - p_\theta}
{(\gamma_{\text{DM}} - \gamma_{\text{BM}})}, \\
\rho_{\text{DM}} &=& -\frac{\gamma_{\text{BM}}(\rho - p_r + p_\theta) - p_\theta}{(\gamma_{\text{DM}} -\gamma_{\text{BM}})}. \nonumber
\end{eqnarray}
As mentioned earlier, $K$ is thus far still free. Throughout this paper we will consider zero radial pressure for mathematical simplicity. In order to describe the collapse of an anisotropic fluid with vanishing radial pressure we set $K$ by requiring that this condition be satisfied by the effective fluid that describes cosmological baryonic and dark matter, i.e. $K = u^\mu v_\nu$ satisfies,
\begin{equation}
K^2 = \frac{-\gamma_{\text{BM}}\rho_{\text{BM}}^2 -\gamma_{\text{DM}}\rho_{\text{DM}}^2 + (1 + \gamma_{\text{BM}}\gamma_{\text{DM}})\rho_{\text{BM}}\rho_{\text{DM}}}{(1 + \gamma_{\text{BM}})(1 + \gamma_{\text{DM}})\rho_{\text{BM}}\rho_{\text{DM}}}.
\end{equation}
For this choice of $K$, the effective anisotropic fluid has zero radial pressure and we have,
\begin{eqnarray} \label{eq:rho_p_eff}
\rho_{\text{BM}} &=& \rho\frac{\gamma_{\text{DM}}(1 + \kappa) - \kappa}
{(\gamma_{\text{DM}} - \gamma_{\text{BM}})}, \\
\rho_{\text{DM}} &=& -\rho\frac{\gamma_{\text{BM}}(1 + \kappa) - \kappa}{(\gamma_{\text{DM}} -\gamma_{\text{BM}})}. \nonumber
\end{eqnarray}
where we have introduced a constitutive equation of state for the effective anisotropic fluid, 
\begin{equation}
\kappa = \frac{p_\theta}{\rho}.
\end{equation}
We can write eq.~(\ref{eq:rho_eff}) as,
\begin{eqnarray}
\label{eq:rho_eff2}
\rho &=& \frac{\rho_{\text{BM}} + \rho_{\text{DM}}}{1 + \kappa}, \\
p_r &=& 0, \nonumber \\
p_\theta &=& p_{\text{BM}} + p_{\text{DM}}. \nonumber
\end{eqnarray}
It can be shown general relativistically \cite{JMN11} that a collapsing anisotropic fluid or isotropic fluid can reach to an equilibrium state in an asymptotic comoving time. 
The JMN-1 spacetime describes the end-state of the collapse of an anisotropic fluid with $p_r = 0$ and $\kappa = k_e$, a constant. Therefore, we can treat this spacetime as describing the equilibrium configuration arising out of the collapse of two perfect fluids that satisfy the criterion discussed above. For baryonic and dark matter collapsing to an equilibrium configuration, the geometry of which is given by the JMN-1 spacetime and we can write eq.~(\ref{eq:rho_eff2}) as,
\begin{eqnarray}
\rho &=& \frac{\rho_{\text{BM}} + \rho_{\text{DM}}}{1 + \kappa_e}, \\
p_r &=& 0, \nonumber \\
p_\theta &=& p_{\text{BM}} + p_{\text{DM}}. \nonumber
\end{eqnarray}
As it was said in the introduction, dark matter is generally thought as pressure-less fluid. However, there are many literature \cite{Bettoni:2012xv}-\cite{Bharadwaj:2003iw}, \cite{Dey:2013yga}, \cite{Dey+15} where the gravitational aspects of dark matter with pressure are investigated. In this paper, during the non-linear collapsing phase of primordial dark matter halo, we consider an anisotropic effective fluid which is seeded by dark matter and baryonic matter. In this model, we can consider the pressure inside dark matter cloud zero or non-zero. For dark matter with zero-pressure, using eq.~(\ref{eq:rho_p_eff}) and eq.~(\ref{eq:rho_eff2}) we get,
\begin{eqnarray}
\rho_{DM}&=&\rho_{BM}\bigg\{\frac{\rho}{\rho_{BM}}(1+\kappa)-1\bigg\},\nonumber\\
p_r&=&0,~ p_{\theta}=p_{BM},
\end{eqnarray}
where we consider $\gamma_{DM}$ is zero to make the dark matter pressureless. In this way, one can show that though effective fluid has anisotropic pressure, one component fluid (dark matter) of this effective fluid can have zero pressure. Therefore, the collapsing effective fluid can be thought of as collapsing cold dark matter (CDM) with baryonic matter. In this paper, using the general relativistic technique of equilibrium, we will discuss how this type of effective anisotropic fluid can terminate into a stable configuration. In the next section, we are going to discuss this general relativistic technique briefly and will show how JMN-1 spacetimes can be formed in asymptotic time.

\section{The JMN-1 Spacetimes} \label{sec:JMN1}
Here we develop an extension of the JMN-1 spacetimes \cite{JMN11} and show that these JMN-1 spacetimes arise very generally, as end states of gravitational collapse processes involving anisotropic fluids with vanishing radial pressure for \textit{arbitrary} initial mass profiles, and not just for an initially homogeneous energy-density profile. As a consequence, the JMN-1 spacetime is stable to perturbations in initial data, i.e. it develops from generic initial data. 

The JMN-1 spacetime forms in the asymptotic large time limit of a collapse that slows down, and describes the geometry in the exterior of a naked singularity. However, at large times, the matter configuration has a centrally peaked density and the cloud has nearly attained equilibrium. Here we approximate the JMN-1 spacetime to describe the geometry outside such a quasi-equilibrated cloud of matter with a large centrally peaked density profile. This is of particular interest here given that we want to study the formation of small-scale structures like galaxies and dark matter halos. In this section, we review important aspects of the JMN-1 spacetimes needed, and in the next section, we utilize these spacetimes to form dark matter halos and galaxy-like structures.

\subsection{Setup of the Collapse Process}
The spacetime geometry of a fluid collapsing under gravity, preserving spherical symmetry throughout its evolution, is given in Lagrangian coordinates  $(t,r,\theta, \phi)$, i.e. in coordinates comoving with the fluid ($u^\mu = e^{-\xi}\delta^\mu_{\ t}$), by the metric,
\begin{equation} \label{eq:GCM}
ds^2 = -e^{2\xi(t,r)}~dt^2 + \frac{R^{\prime 2}(t,r)}{G(t,r)}~dr^2 + R(t,r)^2~d\Omega^2, 
\end{equation}
where, for our purposes, we have rewritten the usual $g_{rr} = e^{2\psi(t,r)}$ in terms of a new function $G(t,r)$ in the above, which is related to the velocity profile of the collapsing fluid at the initial epoch. The $^\prime$ denotes a derivative with respect to $r$, and,
\begin{equation}
d\Omega^2 = d\theta^2 + \sin^2\theta~d\phi^2,
\end{equation}
is the standard metric on a unit two-sphere. We shall consider here the spherically symmetric collapse of an anisotropic fluid for which, in comoving coordinates, the stress-energy tensor becomes diagonal,
\begin{equation} \label{eq:TmunuTransverse}
T^\mu_{\ \nu} = (-\rho, p_r, p_\theta, p_\theta).
\end{equation}
We note here that this stress-energy tensor, corresponding to an anisotropic fluid, will be the effective description of the baryonic and dark matter content of the spacetime in a cosmological setting, as discussed in \S\ref{sec:BMDM}.

In terms of the Misner-Sharp mass function $F(t,r)$ which measures the amount of mass contained within a shell of comoving radius $r$ on the spacelike hypersurface given by $t = const.$ \cite{MisnerSharp64, CahillTaub71}, 
\begin{equation} \label{eq:MSMass}
F(t,r) = R(1 - G(t,r) + e^{-2\xi}\dot{R}^2),
\end{equation}
the governing Einstein equations (EEs) $\mathbb{G}^\mu_{\ \nu} = T^\mu_{\ \nu}$ for the evolution of such a fluid are (see for example \cite{CahillTaub71, JoshiMalafarina11}),
\begin{eqnarray} 
\rho &=& \frac{F^\prime}{R^2 R^\prime}, 
\label{eq:rhoEE} \\
p_r &=& -\frac{\dot{F}}{R^2 \dot{R}}, 
\label{eq:pr} \\ 
\xi^\prime &=& 2\frac{p_\theta - p_r}{\rho + p_r}\frac{R^\prime}{R} - \frac{p_r^\prime}{\rho + p_r}, 
\label{eq:xiPrime} \\
\dot{G} &=& 2 G \dot{R}  \frac{\xi^\prime}{R^\prime}.
\label{eq:Gtr=0}
\end{eqnarray}
In the above, we have used $8\pi G_N = c = 1$. The structure of the EEs is that there are seven dynamical functions denoted here by the set $\{F, \rho, p_r, p_\theta, \xi, G, R\}$ and five equations, requiring two supplementary conditions to close out this system of equations. Here, we shall treat $p_r, p_\theta$ to be the two free functions. After picking these functions, we need only to specify the initial data for the remaining five functions and the EEs evolve the collapse to all later times. 
A discussion on the initial data is given in \S\ref{sec:Initial_Data} below.

Since we want to link equilibrium structure formation with the already existing JMN-1 models, for simplicity, we shall choose the first supplementary condition to be,
\begin{equation}
p_r = 0.
\end{equation}
This has the profound simplification that $\dot{F} = 0$, i.e. $F = F(r)$, implying that it is specified entirely, in this case, simply by specifying the initial data for the collapse. A corollary of this feature is that the mass of the entire cloud is conserved throughout the collapse $F(r_b) = 2 M$. Further, such a metric can always be matched to an exterior Schwarzschild spacetime with a total mass of $M$ at a boundary $r = r_b$ \cite{Israel66}. Also, this allows us to introduce a constitutive equation of state, $\kappa$, from eq. (\ref{eq:xiPrime}) as, 
\begin{equation} \label{eq:kappa}
\kappa=\frac{p_\theta}{\rho} = \frac{R}{2}\frac{\xi^\prime}{R^\prime}.
\end{equation}  
$\kappa$ will henceforth be treated as a proxy for $p_\theta$, the only remaining free function.

The metric function $R(t,r)$ has the significance of being the proper radius of a shell of matter present at comoving radius $r$ and time $t$. Therefore, $R^\prime = 0$ corresponds to the collision of different radial shells of matter, causing what are known as shell-crossing singularities. These types of singularities are gravitationally weak singularities and are removable by a suitable change of coordinates \cite{Clarke93}. Detailed comment on both avoidances of shell-crossing singularities and energy conditions is under preparation and will be reported elsewhere \cite{Dey+18}. In these collapse processes, we simply impose here $R^\prime > 0$ to not have to worry about such shell-crossing singularities arising during our collapse process. Also, we note that all collapse processes considered here satisfy the weak energy condition and strong energy condition. To fulfill these two energy conditions we need,  $\rho + 2p_\theta \geq 0$ and $\rho >0$. That is, we have throughout the collapse $F^\prime > 0$ and $\kappa(t,r) \geq -1/2$.

\subsection{Initial Data} \label{sec:Initial_Data}
After picking the two free functions $p_r, \kappa$, one would naively think that prescribing initial data on some spacelike hypersurface $t = t_i$, denoted by $\Sigma_{t_i}$, comprises of choosing five functions, $\{F_i, \rho_i, \xi_i, G_i, R_i\}$. However, since one requires that the Hamiltonian and momentum constraint equations hold on $\Sigma_{t_i}$ \cite{ADM59}, it becomes evident that the initial data does not comprise of five \textit{independent} functions.

%

Now, we proceed to set the initial data systematically. Since we have already chosen a supplementary condition, $p_r = 0, F$ is time independent. Therefore, once we choose $R_i$, it is clear from eq.~(\ref{eq:rhoEE}) that prescribing $\rho_i$ fixes $F(r)$ throughout the collapse,
\begin{equation} \label{eq:Fr}
F(r) = \int_0^r \rho_i(\tilde{r})\tilde{r}^2~\text{d}\tilde{r}.
\end{equation}
Furthermore, we have also,
\begin{equation}
\xi_i = \int_0^r \frac{2\kappa_i(\tilde{r})}{r}~\text{d}\tilde{r},
\end{equation}
that is, the choice of the free function fixes the initial condition for $\xi$. Further, picking $G_i$ fixes $\dot{R}_i$ from eq.~(\ref{eq:MSMass}) and hence also $\dot{G}_i$ from eq.~(\ref{eq:Gtr=0}). In summary, one needs to just pick independently $\{\rho_i, G_i, R_i\}$, corresponding to the initial density and velocity profile of the matter cloud and the initial scaling respectively.

\subsection{Equilibrium Configurations}
Following \cite{JMN11}, we now set up the conditions for equilibrium to be attained. For any fixed $r$, the equation of motion for the scale factor $R$ can be written in terms of an effective potential as eq.~(\ref{eq:MSMass}),
\begin{equation}
V_{\text{eff}}(t,r) = -\dot{R}^2 = -e^{2\xi}\left(\frac{F}{R} + G - 1\right).
\end{equation}
The condition for collapse to occur is simply given as $\dot{R} < 0$, and for equilibrium to be reached is $\dot{R} = \ddot{R} = 0$. Let us say that until equilibrium is reached, say at some asymptotic time $t_e \! \rightarrow \! \infty$ (see \S3 of \cite{JMN11} for further discussion), the cloud of matter was collapsing, i.e. we have,
\begin{equation}
\dot{R}(t < t_e, r) \! < 0,\ \lim_{t \to t_e} \dot{R}(t, r) = \lim_{t \to t_e} \ddot{R}(t, r) = 0. 
\end{equation}

This implies $V_{\text{eff}}(t_e,r) \! = \! \dot{V}_{\text{eff}}(t_e,r) = 0$ or equivalently,
\begin{equation} \label{eq:GedotGe}
G_e = 1 - \frac{F}{R_e}, \dot{G}_e = \frac{F\dot{R}_e}{R_e^2}, 
\end{equation}
where in the above we have defined the equilibrium metric functions $R_{e} = R(t_e, r), G_e = G(t_e, r)$. Therefore, the requirement of equilibrium configurations to form as end-states of collapse restricts just $G, \dot{G}$ at equilibrium, as given above. Further, we have from eq.~(\ref{eq:Gtr=0}, \ref{eq:kappa}, \ref{eq:GedotGe}) that,
\begin{equation} \label{eq:Re}
R_e = F\left(1 + \frac{1}{4\kappa_e}\right).
\end{equation}
Also,
\begin{equation}
\rho_e = \frac{F^\prime}{R_e^2 R_e^\prime},  \xi_e^\prime = 2\kappa_e \frac{R_e^\prime}{R_e}.
\end{equation}
Note that at equilibrium all dynamics is frozen and all metric functions and physical properties of matter depend simply on $r$.

Since we are still free to choose $\kappa$, for the simplest case when its equilibrium value is a constant, $\kappa_e = k_e$, the metric functions become,
\begin{eqnarray} \label{eq:kappae_ke}
R_e &=& F\left(1 + \frac{1}{4 k_e}\right), R_e^\prime = F^\prime\left(1 + \frac{1}{4 k_e}\right), \\
G_e &=& \frac{1}{1 + 4 k_e}, \xi_e^\prime = 2\kappa_e \frac{F^\prime}{F} = 2\kappa_e \frac{R_e^\prime}{R_e}, \nonumber
\end{eqnarray}
and the equilibrium energy density is,
\begin{equation}
\rho_e = F^{-2}\left(1 + \frac{1}{4 k_e}\right)^{-3},
\end{equation}
where $F = F(r)$ is just given from the initial data in eq.~(\ref{eq:Fr}). We can perform a simple integration of the differential equation of $\xi_e$ in eq.~(\ref{eq:kappae_ke}), to get the expression of $\xi_e$,
\begin{equation} \label{eq:xiEGen}
e^{2\xi_e} = \left(\frac{R}{R_B}\right)^{4k_e}e^{2\xi_B},
\end{equation}
where $R_B \! = \! R_e(r \! = \! r_B), \xi_B \! = \! \xi_e(r \! = \! r_B)$ and the metric is given as,
\begin{equation} \label{eq:Unmatched_JMN1}
ds^2 = -\left(\frac{R_e}{R_B}\right)^{4k_e}e^{2\xi_B} dt^2 + (1+4k_e) dR^2 + R^2 d\Omega^2.
\end{equation}
Due to the freedom in the choice of $\xi_B$, we shall call the above metric the unmatched JMN-1 metric. This terminology will become clear in what follows. Matching this metric on the hypersurface $R_e = R_B$ with the Schwarzschild metric,
\begin{equation}
ds^2 = -\left(1 - \frac{2M}{R_e}\right)dt^2 + \frac{dR_e^2}{\left(1 - \frac{2M}{R_e}\right)} + R_e^2~d\Omega^2,
\end{equation}
yields,
\begin{equation}
e^{2\xi_B} = 1 - \frac{F(R_B)}{R_B} = \frac{1}{1 + 4k_e},
\end{equation}
where we have used the fact that $F(R_B) = 2M$. Now, in terms of a new parameter $\chi$,
\begin{equation} \label{eq:chidef}
\chi = \frac{4k_e}{1 + 4k_e},
\end{equation}
and we recognize that (\ref{eq:Unmatched_JMN1}) when matched with the Schwarzschild metric on its exterior boundary becomes simply the JMN-1 spacetime,
\begin{equation}
ds^2 = -(1-\chi)\left(\frac{R_e}{R_B}\right)^{\frac{\chi}{1-\chi}}~dt^2 + \frac{dR_e^2}{1-\chi} + R_e^2d\Omega^2.
\end{equation}
That is, for any arbitrary initial density profile, in asymptotic time, we obtain the JMN-1 spacetime as the metric for an eventual equilibrium configuration when the collapsing fluid satisfies, $p_r = 0$ and $\kappa(t\rightarrow t_e, r) = k_e$, a constant.


\section{A New Model of Galaxy and Halo Formation} \label{sec:Galaxy_Halo}
In the spherical top-hat collapse model, over-dense regions are described by the closed FLRW metric and initially undergo a phase of expansion, up until a time $t=t_{m}$ when the density of the region reaches approximately $5.55$ times the background density. At this time, these patches start collapsing and these over-dense regions are treated as isolated sub-universes. Within the top-hat collapse model, the spacetime structure exterior to that of the over-dense region is ignored and the comoving radius $r$ ranges from $0$ to $1$. 

In top-hat collapse model, the closed FLRW spacetime radially ranges from $0$ to $1$. Therefore, the overdense region is totally isolated from the background expanding universe. But, to make a more realistic model of structure formation one has to consider background effects. There are series of papers on the spherical gravitational collapse in an expanding background \cite{Mimoso:2009wj}-\cite{Delliou:2013xra}. In these papers, how a separating shell can divide the expanding and collapsing regions, is discussed. One can simplify the model of collapsing matter in expanding background by considering the scale of the primordial dark matter halo very small compare to the Hubble radius of expanding universe. This is a valid assumption, as the radius of Hubble horizon, generally, almost  one thousand times bigger than the radius of dark matter halo. With this assumption, one can prove that the killing vector of external spacetime at the scale of collapsing halo becomes timelike \cite{Meyer:2012nw},\cite{Creminelli:2009mu}. Therefore, one can neglect the expansion of universe at the halo scale. So, in the immediate vicinity of dark matter halo one can consider a static exterior spacetime. The simplest assumption is to consider Schwarzschild spacetime in the immediate vicinity of the overdense region.
Therefore, in this paper, we describe the expanding over-dense regions, as usual, by the closed FLRW spacetime. However, for the above-mentioned reason, we consider the spacetime exterior to this closed FLRW spacetime to be given by the Schwarzschild spacetime, until $t = t_m$.         Till this time, these over-dense regions experience a phase of expansion. Like the top-hat collapse model, here also the  density of the overdense region becomes $5.55$ times the density of the background at turnaround time ($t_m$). The background density is the density of the universe at the cosmological scale, where spatially flat FLRW metric is used to describe the expansion of the homogeneous and isotropic universe. 
\begin{figure}[t]
\includegraphics[scale=0.40]{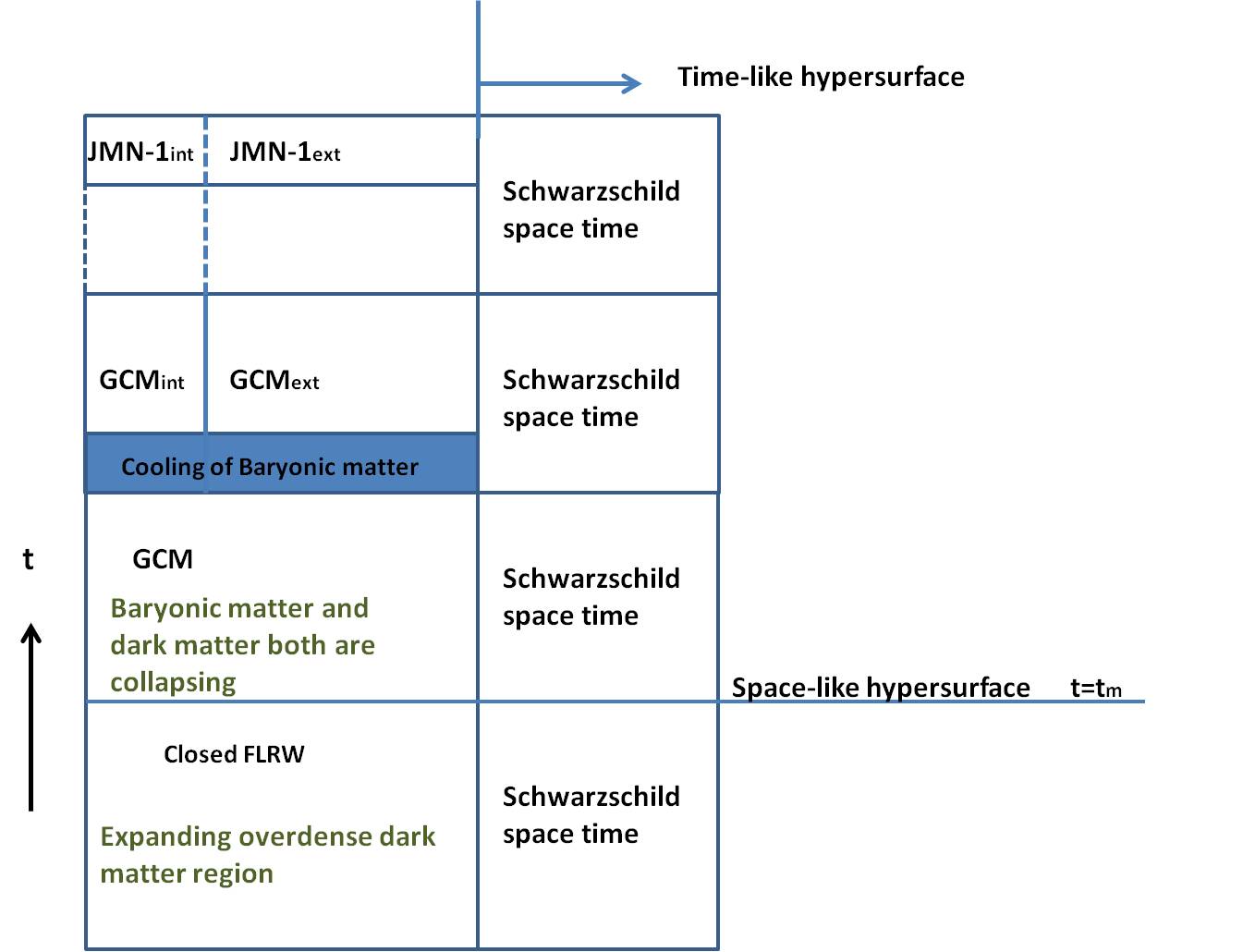}
\caption{ Here, the schematic diagram shows how galactic halo-like structure evolves through different stages when the cooling time of baryonic matter $t_{cool}<t_{dyn}$. }
\label{structure1}
\end{figure}

As we have discussed in subsection (\ref{galform}), when the primordial dark matter halo starts to collapse, the baryonic matter inside the halo also starts collapsing. This contraction in baryonic matter field heats up the baryonic gas and the gas virializes at a virialization temperature. During the expansion period of dark matter halo, one can consider the dark matter as a pressure-less fluid carrying some properties of background dark matter field which is homogeneous, isotropic and pressure-less. 
During the latter part of the expansion period of dark matter halo, the baryonic matter inside the halo region may be thought as pressure-less fluid, as there is no contraction. Consequently, both the dark matter and baryonic matter behave like dust during the expansion phase. Therefore, like the top-hat collapse model, in order to describe the expansion phase of the overdense regions, we use closed FLRW metric. However, as it was mentioned, in our case, during expansion, we construct a spacetime structure which is internally closed FLRW spacetime and externally Schwarzschild spacetime. According to the top-hat collapse model, after the expansion phase, dark-matter halo starts collapsing under its own gravity and virializes to a stable structure. One can show that the general relativistic technique which is discussed in section (\ref{sec:JMN1}), can be thought as the general relativistic analog of Newtonian virialization process \cite{Dey2+18}. Using this technique, one can general relativistically explain how a finite pressure which can be formed inside the collapsing baryonic and dark matter field, can lead to the virialized structures of baryonic matter and dark matter. As it is discussed in the section (\ref{sec:BMDM}), one can consider an effective anisotropic fluid as a combination of two non-interacting isotropic fluids. Therefore, one can explain the effective dynamics of baryonic and dark matter during the contraction phase, by considering an effective anisotropic, collapsing fluid. From the starting point of the contraction phase, one can use a generalized collapsing metric (GCM) to describe the anisotropic fluid collapse. In order to describe the collapsing phase by a GCM, we must ensure the smooth matching of the GCM with the closed FRW spacetime at the time $t=t_m$ on a spacelike hypersurface and  another smooth matching with exterior Schwarzschild metric on a timelike hypersurface.

We know from the previous discussion that baryonic matter cannot retain its virialized structure for a long time. In a finite time, it dissipates its energy and accumulates into the central region of the dark matter halo. Due to this mechanism, the density of baryonic matter in the outer region of the halo becomes much smaller compared to the baryonic  matter density in the central region of the halo. In the central region, the density of baryonic matter can be comparable to or greater than the density of dark matter. In order to describe this mechanism general relativistically, we have to construct a dynamical spacetime structure which is internally GCM$_{int}$ and externally GCM$_{ext}$, where we denote internal and external GCM by GCM$_{int}$ and GCM$_{ext}$ respectively. GCM$_{int}$ is used to describe the further collapse of the central region of the halo, where the baryonic density is very high. On the other hand, GCM$_{ext}$ describes the evolution of the outer region of the halo, where the baryonic matter density is very small compared to the dark matter density. Though the transformation from previous GCM to GCM$_{int}$ and GCM$_{ext}$ is continuous, in this paper we will not attempt to explain this transformation general relativistically. Both the interior and exterior GCM should be matched on some timelike hypersurface. In fig.~(\ref{structure1}), one can see a schematic diagram which shows the different stages of the evolution of halo-like structures as described by our general relativistic model. As we have discussed, in this paper, we will discuss a simpler case where the radial pressures of the general collapsing metrics are zero. Therefore, if we consider that the baryonic matter in the central region of the halo would not further dissipate its energy, then the collapsing GCM$_{int}$ can reach the final equilibrium state in an asymptotic time. On the other hand, as dark matter is dissipation-less, we can always use the general relativistic technique of equilibrium to investigate the asymptotic stable state of the collapsing GCM$_{ext}$. Since we consider zero radial pressure for both the collapsing external and internal GCM, the asymptotic stable spacetime structure will be internally JMN-1$_{int}$ and externally JMN-1$_{ext}$, where the outer boundary of the exterior JMN-1 metric is matched, as usual, with a Schwarzschild metric. In reality, as we know, baryonic matter can further dissipate its energy or it can have a large amount of angular momentum during the collapse of the central part of the halo. Therefore, it would be a very much idealistic situation to get JMN-1$_{int}$ as the final state of the gravitational collapse of the central part of the halo region. Nevertheless, our model can be thought of as a toy model of galactic halo formation which can be generalized in the future to describe more realistic situations. In the next subsections, we will see this general relativistic model of structure formation can give some useful predictions.

\begin{figure}[t!]
\includegraphics[width=8cm,height=8cm]{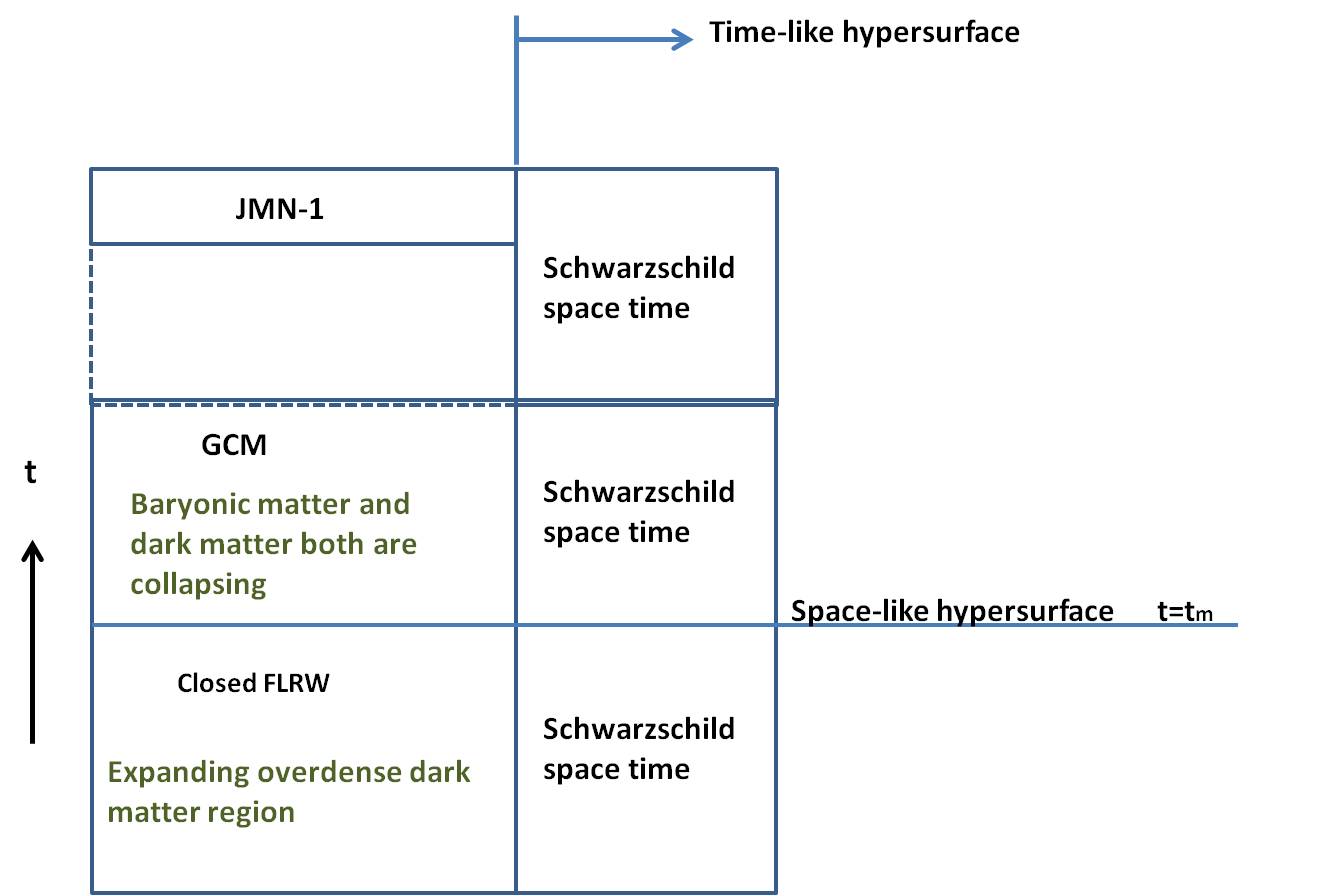}
\caption{Here, the schematic diagram shows how galactic halo-like structure evolves through different stages when the cooling time of baryonic matter $t_{cool}>t_{dyn}$. }
\label{structure2}
\end{figure}

As we know, for the case when $t_{cool}>t_{dyn}$, baryonic gas will be pressure supported and will collapse quasi-statically \cite{White:1977jf}. For dark matter one can always consider $t_{cool}>t_{dyn}$, as the dark matter is dissipationless. One can consider $t_{cool}\geq H^{-1}_{present}$ for dark matter, where $H_{present}$ is the present value of Hubble parameter. Therefore, dark matter collapse should be always quasi-static and pressure supported. For that reason, in the previous case, where $t_{cool}<t_{dyn}$ for baryonic matter, we have considered pressure supported, quasi-static dark-matter collapse. Therefore, in the present case, both the baryonic and dark matter collapse quasi-statically. One can model these two component fluids collapse as the collapse of an anisotropic fluid collapse. This quasi-static collapse process of an anisotropic fluid can be described by a GCM, as it is shown in  fig.~(\ref{structure2}). Since here also for simplicity, we have chosen zero radial pressure, the asymptotic spacetime will be JMN-1 spacetime. 

In the subsection (\ref{galform}), we have discussed another model where a major portion of baryonic matter dissipates its energy before it reaches to the virialized state and accumulates to the central region of the halo where it collapses further. This model can also be described by fig.~(\ref{structure1}), considering the time span of GCM is much lesser than the previous case.  

\begin{figure*}[t!]
\centering
\includegraphics[width=12cm,height=10cm]{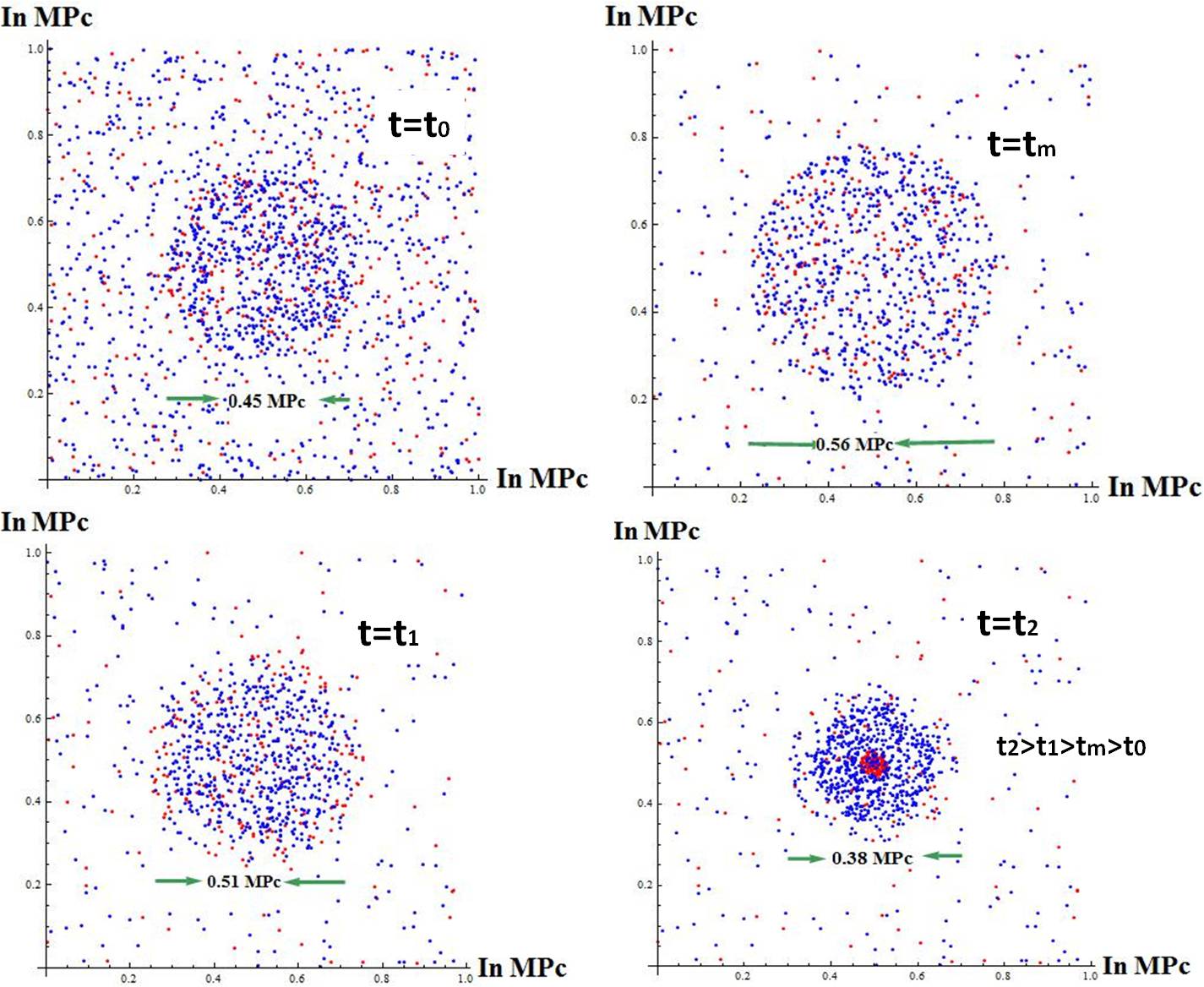}
\caption{It is a cartoon picture where the evolution of structures is shown. The density of blue dots and red dots corresponds to the density of dark matter and baryonic matter respectively. The top left corner diagram is showing the starting point (time $t=t_0$)  of the non-linear expansion of overdense region. The top right corner diagram is showing the situation (time $t=t_m$) when the overdense region stops expanding and starts collapsing. The bottom left corner diagram is indicating some inhomogeneity in density during the collapse, whereas bottom right corner diagram is showing the accumulation of baryonic matter in the central region of the halo.     }
\label{structure3}
\end{figure*}

In fig.~(\ref{structure3}), it is diagrammatically shown how the baryonic and dark matter structures evolve with time. In that figure, dark matter density and baryonic matter density correspond to the density of blue and red dots respectively. At time  $t=t_0$ the overdense region of dark matter starts expanding non-linearly and we consider at that time its diameter is $0.45$ MPc. At time  $t=t_m$ it reaches to the turnaround point where from the overdense region starts collapsing due to its own gravity. We consider that the turnaround diameter of the overdense patch is $0.56$ MPc. Baryonic matter also collapses with dark matter and after some time, by dissipating energy, it accumulates at the central region of the halo. At time $t=t_2$, the central part of halo, where the baryonic density is very high, starts collapsing again. The time $t=t_1$ is the intermediate time in between $t_m$ and $t_2$. One can see some inhomogeneity form in the dark matter field during the collapse. Since the background is expanding, in fig.~(\ref{structure3}), one can see that the background density decreases with time. As in that figure the turnaround diameter is         $0.56$ MPc, the virialized diameter should be around $0.28$ MPc. However, one can see that the baryonic matter starts dissipating its energy before it reaches the virialized state. Therefore, fig.~(\ref{structure3}) describes the structure formation process, as it is stated by the second model which is discussed in subsection (\ref{galform}).  

It is well-known that in GR to see if two metrics match smoothly, i.e. without any matter shell at the junction of the two metrics. For this, one should match both the induced metrics and the extrinsic curvatures on that hypersurface smoothly to obtain the Israel junction conditions \cite{Israel66}. These junction conditions can give the initial conditions for collapse, which is very essential information for predicting the final equilibrium state of collapse. In the following, we 
will discuss the required junction conditions for our model and their consequences. According to the fig.~(\ref{structure1}), we need to match: 1) expanding closed FLRW metric with exterior Schwarzschild metric on a timelike hypersurface,  2) expanding closed FLRW metric with a GCM on a spacelike hypersurface, 3) internal GCM with external Schwarzschild metric on a timelike hypersurface, 4) Internal GCM with external GCM on a timelike hypersurface, 5) internal JMN-1 with external JMN-1 on a timelike hypersurface. 

\subsection{Matching of an Interior Closed FLRW spacetime with an Exterior Schwarzschild spacetime}
Our first task is to match the interior closed FLRW spacetime with an exterior Schwarzschild spacetime, whose metrics are given as,
\begin{eqnarray}
ds^2_{\text{cFLRW}} &=& -d\tau^2+a^2(\tau)\left(\frac{dr^2}{1-r^2}+r^2d\Omega^2\right), \\
ds^2_{\text{Schw}} &=& -h(l)dt^2+h(l)^{-1}dl^2+l^2d\Omega^2,
\end{eqnarray}
where $h(l)=1-\frac{2M}{l}$. The above two metrics can be matched on a timelike hypersurface given by $r=r_0$, where $0 < r_0 < 1$, which we shall denote by $\Sigma$. Induced co-ordinates on $\Sigma$ are $y^a=(\tau,\theta, \phi)$. From the exterior Schwarzschild spacetime, $\Sigma$ is described by the parametric equations $l = L(\tau)$ and $t = T(\tau)$ and the induced metric derived from the Schwarzschild metric is,
\begin{equation}
ds_{\Sigma, \text{Schw}}^2 = -(H\dot{T}^2 - H^{-1}\dot{L}^2)d\tau^2 + L^2(\tau)d\Omega^2,
\end{equation}
where $H=1-\frac{2M}{L}$. The induced metric on $\Sigma$ obtained from the interior closed FLRW metric is,
\begin{equation}
ds_{\Sigma, \text{cFLRW}}^2 = -d\tau^2+a^2(\tau)r_{0}^2d\Omega^2.
\end{equation}
The condition that the metrics on $\Sigma$ induced from both the interior closed FLRW and the exterior Schwarzschild metrics match smoothly yields,
\begin{equation}
L(\tau)=a(\tau)r_0, (H\dot{T}^2-H^{-1}\dot{L}^2)=1.
\end{equation}
The extrinsic curvature of the hypersurface is expressed as,
\begin{equation} \label{eq:extrinsic_curvature}
K_{ab}=n_{\alpha ; \beta}e^{\alpha}_{a}e^{\beta}_{b}
\end{equation}
where $n_{\alpha ; \beta}$ is the covariant derivative of $n_{\alpha}$ which is a normal to the matching hypersurface and  $e^{\alpha}_{a} \equiv \frac{\partial x^{\alpha}}{\partial y^a}$, where $y^a$ is the induced coordinate on the hypersurface.
The interior and exterior normals to $\Sigma$ are given respectively by,
\begin{eqnarray}
n^{\alpha}_{\text{cFLRW}} &=& \left\{0,\frac{a(\tau)}{\sqrt{1-r^2}},0,0\right\}, \\
n^{\alpha}_{\text{Schw}} &=&\{-\dot{L},\dot{T},0,0\}
\end{eqnarray}
The extrinsic curvature of $\Sigma$ obtained from the interior closed FLRW spacetime is then given by,
\begin{eqnarray}
(K_{tt})_{\text{cFLRW}} &=& 0, \\
(K_{\theta\theta})_{\text{cFLRW}}  &=& a(\tau)r\sqrt{1-r^2}, \nonumber
\end{eqnarray}
and the same from the exterior Schwarzschild spacetime is given as,
\begin{eqnarray}
(K_{tt})_{\text{Schw}} &=& -\dot{L}^{-1}\frac{d}{d\tau}(H\dot{T}), \\
(K_{\theta\theta})_{\text{Schw}} &=& LH\dot{T}. \nonumber
\end{eqnarray}
Then, from the condition that the extrinsic curvatures for $\Sigma$ obtained from both the interior and exterior metric match smoothly, we obtain,
\begin{equation}
(H\dot{T})^2=\dot{L}^2+H=(1-r_0^2),
\end{equation}
which is simply the energy conservation equation. Since the external Schwarzschild solution is a vacuum solution, this statement of energy conservation implies that the total mass inside the collapsing system is constant. 

Before the time $t=t_m$, the collapsing system should maintain the above junction conditions, and at this time the closed FLRW spacetime must then be matched smoothly with a general collapsing metric (GCM) on a spacelike hypersurface. Further, at the same time, GCM must also be smoothly matched with an exterior Schwarzschild spacetime on a timelike hypersurface.

\subsection{Matching of Closed FLRW spacetime with a General Collapsing Metric (GCM) on a Spacelike Hypersurface}
Here we discuss the matching of the closed FLRW metric with that of a general collapsing metric (GCM), the following two spacetimes, at the time $t = t_m$, on a spacelike hypersurface,
\begin{eqnarray}
ds^2_{\text{cFLRW}} &=& - d\tau^2 + {a^2(\tau)\over 1- r^2 }dr^2 + a^2(\tau) r^2 d\Omega^2, \\
ds^2_{\text{GCM}} &=& - e^{2\nu(r,t)} d\tau^2 + {R'^2\over G(r,\tau)}dr^2 + R^2(r,\tau) d\Omega^2. \nonumber
\end{eqnarray} 
At the time $t = t_m$, the scale factor of the closed FLRW $a(t)$ reaches its maximum value $a_m$, and therefore we have $\dot{a}(t_m)=0$. Therefore, on this spacelike hypersurface $\Sigma$, we can write the induced metrics as,
\begin{eqnarray}
ds^2_{\Sigma, \text{cFLRW}} &=& \frac{a_m^2 dr^2}{1-r^2} + a_m^2r^2 d\Omega^2, \\
ds^2_{\Sigma, \text{GCM}} &=& \frac{R^{\prime 2}(r,t_{\rm max})dr^2}{G(r,t_m)}+R^2(r,t_{\rm max})d\Omega^2
\end{eqnarray}
For closed FLRW metric, the extrinsic curvature can be written as,
\begin{eqnarray}
(K_{11})_{\text{FLRW}}&=&\frac{a(t)\dot{a}(t)}{1-r^2}\,\, ,\nonumber\\
(K_{22})_{\text{FLRW}}&=&\frac{(K_{33})_{\text{FLRW}}}{\sin^2\theta}=r^2a(t)\dot{a}(t)\,\, .
\end{eqnarray}
Since $\dot{a}(t_m)=0$, all the components of the extrinsic curvature of FLRW metric become zero on the spacelike hypersurface $\tau=t_{m}$.
In order to match the collapsing metric to the FLRW metric at $\tau=t_{m}$, we require
that the components of the extrinsic curvature of the former also be set to zero at this time. The extrinsic curvature of GCM is given as,
\begin{widetext}
\begin{eqnarray}
(K_{11})_{\text{GCM}} &=& \frac{e^{-\nu (r,\tau)}}{2 G(r,\tau)^2}\left(r f'(r,\tau)+f(r,\tau)\right)\left(\left(r f'(r,\tau)+f(r,\tau)\right) {\dot G}(r,\tau)-2 \left({\dot f}(r,\tau)+r{\dot f}'(r,\tau)\right) G(r,\tau)\right)\nonumber\\
(K_{22})_{\text{GCM}} &=& -r^4 f(r,\tau) {\dot f(r,\tau)} \left(e^{-\nu (r,\tau)}\right) = \frac{(K_{33})_{\text{GCM}}}{\sin^2\theta},
\end{eqnarray}
\end{widetext}
where we have introduced, $R=rf(r,\tau)$ using the scaling degree of freedom. Now, from the matching of the metric on both sides it is clear that the initial value of $\nu(r,\tau)$, $f(r,\tau)$ and $G(r,\tau)$ at $t_m$ should be,
\begin{eqnarray}
\nu(r,t_m)=0, f(r,t_m) = a_m, G(r,t_m) = 1-r^2.
\label{nmatch}
\end{eqnarray}
From the extrinsic curvature matching one obtains,
\begin{eqnarray}
 f(r, t_{m}) \dot{G}(r,t_{m}) = 2 \dot{f}(r,t_{m})
 G(r,t_{m}) = 0\,,
\label{cons1}
\end{eqnarray}
which implies that $\dot{G}(r,\tau_{m})=0$. So, from the smooth matching of a closed FLRW metric and the GCM on the spacelike hypersurface, we get following conditions,
\begin{eqnarray} \label{Gmatch}
\nu(r,t_m) &=& 0, f(r,t_m) = a_m, G(r,t_m) = 1-r^2, \nonumber\\ 
\dot{G}(r,t_m) &=& 0, \dot{f}(r,t_m)=0.
\end{eqnarray}
In our case, since the radial pressure is zero, we have $\dot{F}=0$ and consequently, the form of $F$ is obtained once and for all if we know its form at $\tau=t_{m}$. From matching conditions we have,
\begin{eqnarray} \label{msm}
\mathbb{F}(r) &=& rf(r,t_{m})\left[1- G(r,t_{m}) +
r^2 e^{-2\nu(r,t_{m})} \dot{f}^2(r,t_{m})\right] \nonumber \\
&=& a_{m} r^3,
\end{eqnarray}
which is a regular function throughout the contracting phase. Now at the time $t = t_m$, and throughout the collapse, one should match GCM with external schwarzschild and we discuss this in the next subsection.

\subsection{Matching of  Internal GCM with External Schwarzschild spacetime on a Timelike Hypersurface}
We will now match the internal GCM with an external Schwarzschild spacetime,
\begin{eqnarray}
ds_{\text{GCM}}^2 &=& - e^{2\nu(r,\tau)} d\tau^2 + {R'^2\over G(r,\tau)}dr^2 + R^2(r,\tau) d\Omega^2 \nonumber \\
ds^2_{\text{Schw}} &=& -h(l)dt^2+h(l)^{-1}dl^2+l^2d\Omega^2,
\end{eqnarray}
We match these two metrics on a timelike hypersurface $\Sigma$, $r=r_0$, where $0\leq r_0<1$ and $y^a=(\tau,\theta, \phi)$ are induced coordinates on it. The induced metric on $\Sigma$ obtained from interior GCM is,
\begin{equation}
ds_{\Sigma, \text{GCM}}^2=- e^{2\nu(r_0,\tau)} d\tau^2+\mathbb{R}^2(r_0,\tau)d\Omega^2,
\end{equation}
and the induced metric from the exterior Schwarzshild is given as,
\begin{equation}
ds_{\Sigma, \text{Schw}}^2=-(H\dot{T}^2-H^{-1}\dot{L}^2)d\tau^2+L^2(\tau)d\Omega^2,
\end{equation}
where $H=1-\frac{2M}{L}$. The condition for smooth matching of these induced metric gives,
\begin{equation}
L(\tau)=\mathbb{R}(r_0,\tau),\ \  (H\dot{T}^2-H^{-1}\dot{L}^2)= e^{2\nu(r_0,\tau)}.
\end{equation}
Using the same previous procedure one can derive the expression of extrinsic curvature, and from their matching, one obtains
\begin{eqnarray}
(H\dot{T})^2=\left[\dot{L}^2+H e^{2\nu(r_0,\tau)}\right]=G(r_0,\tau).
\end{eqnarray}
From the matching of the temporal part of extrinsic curvature of both sides we can see that the term $H\dot{T}$ is time independent and consequently from the above equation $G(r_0,\tau)$ becomes time independent. So the matching condition implies that,
\begin{equation}
\frac{d}{d\tau}G(r_0,\tau)=0.
\end{equation}
One can recall that we have derived the above condition in the previous subsection eq.~(\ref{Gmatch}) where we match closed FRW and GCM on a spacelike hypersurface.
From the previous subsection we get,
\begin{eqnarray}
\nu(r,t_m) &=& 0, f(r,t_m) = a_m, G(r,t_m) = 1-r^2, \\
\dot{G}(r,t_m) &=& 0, \dot{f}(r,t_m)=0.
\end{eqnarray}
One can check that the above conditions fulfill the requirement of smooth matching of internal GCM with the external Schwarzschild spacetime on a timelike hypersurface at time $t_m$. But these conditions are not sufficient for smooth matching of internal GCM and external Schwarzschild spacetime throughout the collapse.

The condition $\frac{d}{d\tau}G(r_0,\tau)=0
$ implies that $G(r,\tau)$ is independent of time at the boundary ($r=r_0$) throughout the collapse. Now, we know that at equilibrium GCM becomes a JMN-1 spacetime and consequently $G$ becomes, 
\begin{equation}
G(r,\tau\rightarrow t_e)= 1-\chi
\end{equation}
At the boundary the value of $G$ is time independent, and the value of $\chi_2$ is fixed from the time when the system started collapsing. This implies,
\begin{equation}
\chi=r_0^2.
\end{equation}
The above relation indicates a very interesting fact of this total collapsing history. The above relation makes a connection between the parameters of primordial structure and the parameters of present structure. So from the value of parameters of present structure one can understand the physics of primordial over-dense regions. One can derive another consequence of above results,
\begin{equation}
\dot{G}(r_0,\tau)=\left(\frac{2\dot{R}G}{R^{\prime}}\right)_{r=r_0}\nu^{\prime}(r,\tau)|_{r=r_0}=0\, .
\end{equation}
The above equation implies that $\nu(r,\tau)$ has extremum value at $r=r_0$ throughout the collapse. Now one can check that this condition is also fulfilled when the system reaches to its equilibrium state. We know that from previous calculations, 
\begin{equation}
 e^{2\nu(r,\tau\rightarrow t_e)}=(1-\chi)\left(\frac{R}{R_b}\right)^{\frac{\chi}{1-\chi}}\, .
\end{equation}
Now we also know that $R$ is monotonically increasing as function of comoving radius $r$ and $r_0$ is the maximum value of $r$. So the $g_{00}$ component of JMN-1 metric has a maximum value at $r=r_0$.

As we have discussed before, baryonic matter cannot retain its virialized state for a long time. It eventually radiates its energy and accumulates in the central area of dark matter halo. On the other hand, dark matter retains its virialized state and collapses quasi-statically. We describe the central baryonic matter dominant area by $GCM_{int}$ spacetime and remaining dark matter dominant area by $GCM_{ext}$ spacetime. As we mentioned before, in this paper, we do not discuss the transition from $GCM$ to $GCM_{ext}$ and $GCM_{int}$ using the general relativistic technique. As in the previous cases, here also we need to match $GCM_{ext}$ and $GCM_{int}$ at a timelike hypersurface.

\subsection{Matching of GCM$_{int}$ and GCM$_{ext}$ at Comoving Radius $r_{B1}$ }
We need to match following two spacetimes,
\begin{eqnarray}
ds_{\text{GCM}_{int}}^2 &=& - e^{2\nu_{int}(r,\tau)} d\tau^2 + {R'^2\over G_{int}(r,\tau)}dr^2 + R^2(r,\tau) d\Omega^2\nonumber\\
ds_{\text{GCM}_{ext}}^2 &=& - e^{2\nu_{ext}(r,\tau)} d\tilde{\tau}^2 + {\tilde{R}'^2\over G_{ext}(r,\tau)}dr^2 + \tilde{R}^2(r,\tau) d\Omega^2. \nonumber
\end{eqnarray}
On a timelike hypersurface the two induced metrics are,
\begin{eqnarray}
ds^2_{\Sigma, {\text{GCM}_{int}}}=- e^{2\nu_{int}(r_{b1},\tau)} d\tau^2+ R^2(r_{b1},\tau) d\Omega^2 \\
ds^2_{\Sigma, {\text{GCM}_{ext}}} =- e^{2\nu_{ext}(r_{b1},\tau)} d\tau^2+ \tilde{R}^2(r_{b1},\tau) d\Omega^2.
\end{eqnarray}
From metric matching one gets,
\begin{eqnarray} \label{gmatch}
\nu_{int}(r_{b1},\tau) = \nu_{ext}(r_{b1},\tau), R(r_{b1},\tau)=\tilde{R}(r_{b1},\tau),
\end{eqnarray}
and the extrinsic curvature from both sides can be written as,
\begin{eqnarray}
(K_{00})_{\text{GCM}_{int}} &=& 0, \\
(K_{22})_{\text{GCM}_{int}} &=& R(r,\tau)\sqrt{G_{int}(r,\tau)}\, ,\\
(K_{00})_{\text{GCM}_{ext}} &=& 0, \nonumber\\
(K_{22})_{\text{GCM}_{ext}} &=& \tilde{R}(r,\tau)\sqrt{G_{ext}(r,\tau)}.
\end{eqnarray}
From the matching of extrinsic curvatures one gets,
\begin{equation} 
R(r,\tau)\sqrt{G_{int}(r,\tau)}=\tilde{R}(r,\tau)\sqrt{G_{ext}(r,\tau)}
\end{equation}
At the junction the above condition boils down to,
\begin{equation}
G_{int}(r_{b1},\tau)=G_{ext}(r_{b1},\tau).
\label{Gmatch}
\end{equation}
Now for both dynamic spacetimes we can write Misner-Sharp mass term at $r=r_{b1}$,
\begin{eqnarray}
&& F_{int}(r_{b1}) = \\ 
&& R(r_{b1},\tau)\left(1-G_{int}(r_{b1},\tau)+e^{-2\nu_{int}(r_{b1},\tau)}\dot{R}^2(r_{b1},\tau)\right), \nonumber\\
&& F_{ext}(r_{b1}) = \nonumber\\ 
&&\tilde{R}(r_{b1},\tau)\left(1-G_{ext}(r_{b1},\tau)+e^{-2\nu_{ext}(r_{b1},\tau)}\dot{\tilde{R}}^2(r_{b1},\tau)\right). \nonumber
\label{misner}
\end{eqnarray}
Here we consider that for both dynamic spacetimes radial pressure $P_r=0$, and therefore the Misner-Sharp mass terms are time independent for both spacetimes.
From eq.~(\ref{gmatch}),(\ref{Gmatch}),(\ref{misner}) one can get the following result,
\begin{equation}
F_{int}(r_{b1})=F_{ext}(r_{b1}),
\end{equation}
which should be maintained throughout the collapse for smooth matching of GCM$_{int}$ and GCM$_{ext}$. Now, for a simplistic case, if we ignore angular momentum and further radiation  of baryonic matter then it would collapse quasi-statically and would terminate into static JMN-1 spacetime. On the other hand, outside of this central region, dark matter also collapses quasi-statically and forms exterior JMN-1 spacetime. Therefore, we need to match the interior and exterior JMN-1 spacetime at a timelike hypersurface.

\subsection{The Interior \& Exterior JMN-1 Spacetimes} \label{sec:Int_Ext}
We use the $(T, R)$ coordinates in the region described by the interior JMN-1 metric, $(\tilde{T}, \tilde{R})$ for the exterior JMN-1 and outermost Schwarzchild metrics. Also, $0 \leq R \leq R(r_{b1}), \tilde{R}(r_{b1}) \leq \tilde{R} < \infty$ and the exterior JMN-1 metric  is matched to the outermost Schwarzschild metric on the surface $\tilde{R} = \tilde{R}_{b2} \equiv \tilde{R}(r_{b2})$. The three metrics are,
\begin{eqnarray}
ds^2_{\text{int}}&=& -e^{2\xi(R_{b1})}\left(\frac{R}{R_{b1}}\right)^{\frac{\chi_{int}}{1-\chi_{int}}}dT^2 \! + \! \frac{dR^2}{1-\chi_{int}} \! + \! R^2d\Omega^2, \nonumber \\
ds^2_{\text{ext}}&=& -(1-\chi_{ext})\left(\frac{\tilde{R}}{\tilde{R}_{b2}}\right)^{\frac{\chi_{ext}}{1-\chi_{ext}}}d\tilde{T}^2 \! + \! \frac{d\tilde{R}^2}{1-\chi_{ext}} \! + \! \tilde{R}^2 d\Omega^2, \nonumber \\
ds^2_{\text{Schw}}&=&-\left(1 - \frac{\chi_{ext} \tilde{R}_{b2}}{\tilde{R}}\right)d\tilde{T}^2 \! + \! \frac{d\tilde{R}^2}{\left(1 - \frac{\chi_{ext} \tilde{R}_{b2}}{\tilde{R}}\right)} \! + \! \tilde{R}^2d\Omega^2. \nonumber 
\end{eqnarray}
Now, we proceed to study metric and extrinsic curvature matching on the hypersurface $r = r_{b1}$.

At $R = R_{b1}$, in JMN$-1_{\text{int}}$, we have,
\begin{equation}
ds^2_{\text{int}} = -e^{2\xi(R_{b1})}~dT^2 + R_{b1}^2 d\Omega^2. 
\end{equation}
And for $\tilde{R} = \tilde{R}_{b1}$, in JMN$-1_{\text{ext}}$, we have
\begin{equation}
ds^2_2 = -(1-\chi_{ext})\left(\frac{\tilde{R}_{b1}}{\tilde{R}_{b2}}\right)^{\frac{\chi_{ext}}{1-\chi_{ext}}}~dt^2 + \tilde{R}_{b1}^2 d\Omega^2. \nonumber 
\end{equation}
From the matching of azimuthal parts of the above two metrics we obtain,
\begin{equation}
R(r_{b1})=\tilde{R}(r_{b1})
\end{equation}
Further, from the matching of temporal parts of two metrics one gets,
\begin{equation}
e^{2\xi_e(r_{b1})} = (1-\chi_{ext})\left(\frac{\tilde{R}_{b1}}{\tilde{R}_{b2}}\right)^{\frac{\chi_{ext}}{1-\chi_{ext}}},
\end{equation}
Therefore, from induced metric matching, we can write,
\begin{eqnarray}
ds^2_{\text{int}} &=& -(1-\chi_{ext})\left(\frac{\tilde{R}_{b1}}{\tilde{R}_{b2}}\right)^{\frac{\chi_{ext}}{1-\chi_{ext}}}\left(\frac{R}{R_{b1}}\right)^{\frac{\chi_{int}}{1-\chi_{int}}}dt^2 \nonumber \\
&& \quad\quad + \frac{dR^2}{1-\chi_{int}} + R^2 d\Omega^2, \\
ds^2_{\text{ext}} &=& -(1-\chi_{ext})\left(\frac{\tilde{R}}{\tilde{R}_{b2}}\right)^{\frac{\chi_{ext}}{1-\chi_{ext}}}dt^2 \! + \! \frac{d\tilde{R}^2}{1-\chi_{ext}} \! + \! \tilde{R}^2 d\Omega^2. \nonumber
\end{eqnarray}
The normal $n^\alpha$ to the matching hypersurface, $r = r_{b1}$ can be written as, 
\begin{equation} \label{eq:normal}
n^{\alpha} = \{0,\sqrt{g^{rr}},0,0\},
\end{equation}
where induced coordinates $y^a\equiv \lbrace t,\theta,\phi\rbrace$. For smooth matching we need $(K_{ab})_{\text{int}} = (K_{ab})_{\text{ext}}$ at the matching radius $r=r_{b1}$. From (\ref{eq:normal}, \ref{eq:extrinsic_curvature}), we obtain the non-zero components of $K_{ab}$ to be,
\begin{eqnarray}
K_{tt}&=&-\Gamma^{r}_{tt}\sqrt{g_{rr}}\nonumber\\
K_{\theta\theta}&=&K_{\phi\phi}=-\Gamma^{r}_{\theta\theta}\sqrt{g_{rr}}
\end{eqnarray}
For the interior metric, we have
\begin{eqnarray}
(K_{tt})_{\text{int}} &=&-\frac{\left(\frac{R}{R_{b1}}\right)^{\frac{\chi_{int}}{1-\chi_{int}}}\left(\frac{\tilde{R}_{b1}}{\tilde{R}_{b2}}\right)^{\frac{\chi_{ext}}{1-\chi_{ext}}}\chi_{int}(1-\chi_{ext})}{2R\sqrt{1-\chi_{int}}}, \nonumber\\
(K_{\theta\theta})_{\text{int}} &=&-\sqrt{R(1-\chi_{int})}, 
\end{eqnarray}
and for the exterior metric, these are given as,
\begin{eqnarray}
(K_{tt})_{\text{ext}} &=& -\frac{\left(\frac{\tilde{R}}{\tilde{R}_{b2}}\right)^{\frac{\chi_{ext}}{1-\chi_{ext}}}\chi_{ext}\sqrt{1-\chi_{ext}}}{2\tilde{R}}\nonumber\\
(K_{\theta\theta})_{\text{ext}} &=& -\sqrt{\tilde{R}(1-\chi_{ext})}
\end{eqnarray}
From the requirement of smooth matching of extrinsic curvatures of both regions at $r=r_{b1}$, we get the condition that $\chi_{int}=\chi_{ext}$, i.e., 
\begin{equation}
\frac{F(r)}{R}=\frac{\tilde{F}(r)}{\tilde{R}},
\end{equation}
where $\tilde{F}$ is the Misner-Sharp mass function of the fluid that forms the exterior JMN-1 spacetime. Further, since we have $R(r_{b1})=\tilde{R}(r_{b1})$, we have also
\begin{equation}
F(r_{b1})=\tilde{F}(r_{b1})\,\, .
\end{equation}
This is similar to our previous result which was obtained from the matching of the GCM$_{\text{int}}$ and GCM$_{\text{ext}}$ spacetimes. With the simplistic assumption that the baryonic matter, at the central region of halo, will not further dissipate its energy, we can say that one can model a galactic halo-like structure by above mentioned general relativistic model.   
\bigskip

\section{Discussion and Conclusions} \label{sec:Discussion}
We have shown here is that the formation of small scale equilibrium structures such as dark matter halos and galaxies can be approached within the framework of full general relativity. The important and strong simplifications we introduced here are:
\begin{itemize}
\item In our model, we assume 
eventually that $P_r=0$, which is the term interpreted as the radial pressure in the model. However, we have shown that an anisotropic fluid with zero radial pressure can be interpreted as a combination of two perfect fluids. In this paper the two perfect fluids are baryonic and dark matter.
\item We do not consider any further radiation of energy of baryonic matter after they accumulate at the central part of dark matter halo.
\item We neglected the angular momentum of the whole collapsing system. Therefore, for a system with very high angular momentum, our model would not be applicable. 
\end{itemize}
This is done mainly for the sake of simplicity and transparency. In fact, the results do generalize to the case when the radial pressures are non-vanishing, though the formalism gets more complicated. One can also investigate the final outcomes of gravitational collapse of baryonic matter concentrated around the center, considering further radiation of energy from the baryonic matter sector. We plan to discuss these results separately elsewhere.
While we have mainly presented here the key formalism, the next important issue would be to consider the physical applications of these results.


\end{document}